# Manipulating the Optical Response of TaIrTe$_4$ Heterostructures through Band Alignment Strategy

Longfei Guo[1], Shaowen Xu[2], Qilong Cui[1], Qingmin Hu[3], Ruixue Li[1], Gaofeng Xu[1], Fanhao Jia[2†], and Yuan Li[1]*

[1]*Department of Physics, Hangzhou Dianzi University, Hangzhou 310018, China*
[2]*School of Physics and Optoelectronic Engineering, Hangzhou Institute for Advanced Study, University of Chinese Academy of Sciences, Hangzhou 310024, China*
[3]*NEST Lab, Department of Chemistry, College of Science, Shanghai University, Shanghai 200444, China*
†Emails: fanhaojia@shu.edu.cn
*Emails: liyuan@hdu.edu

## Abstract

Weyl semimetals, such as TaIrTe$_4$, characterized by their unique band structures and exotic transport phenomena, have become a central focus in modern electronics. Despite extensive research, a systematic understanding of the impact of heterogeneous integration on the electronic and optical properties of TaIrTe$_4$ device remains elusive. We have carried out density functional theory combined with nonequilibrium Green's function formalism calculations for TaIrTe$_4$/WTe$_2$, TaIrTe$_4$/MoTe$_2$ and TaIrTe$_4$/*h*-BN heterostructures, aiming to understand the manipulation of photoresponse through various band alignment strategies. The underlying impacts of interlayer interactions, charge transfer and build-in electric field on the electronic properties are carefully investigated. We design a dual-probe photodetector device to understand the overall photoresponse enhancement of the heterogeneous integration by decomposing into the specific strain, interlayer transition, band overlap and symmetry lowering mechanics. These van der Waals integrations provide an ideal platform for studying band alignment physics in self-powered optoelectronic devices.

# I. Introduction

TaIrTe4, a ternary transition-metal dichalcogenides (TMDs) with a non-centrosymmetric orthorhombic structure, has garnered significant interest in the field of optoelectronics due to its intriguing electronic properties [1,2] and potential applications in the self-powered broadband photodetection [3-5] and novel optoelectronic devices [6,7]. It is classified as a type-II Weyl semimetal (WSM) and exhibits only two pairs of Weyl points in its first Brillouin zone the minimal nonzero number of Weyl points allowed for a time-reversal invariant WSMs [8]. Moreover, it possesses the longest Fermi arcs among known WSMs [9]. Experimental studies have revealed intriguing behaviors of TaIrTe4, such as room-temperature nonlinear Hall effect [10], the strongest in-plane electrical anisotropy among all known electrically anisotropic materials, a high carrier density comparable to metals like copper and silver [11], and high photoresponsivity [3]. Its quasi-two-dimensional nature also facilitates enhanced light-matter interactions and enables tunability through integrating with van der Waals (vdW) heterostructures.

The strategy of band alignment is central to vdW integration [12,13], involving precise control of the band structure at the interfaces of different layered materials. Its concept is rooted in understanding how the valence and conduction band edges of various materials interact when in close proximity [14]. Extensively applied and studied in semiconductor researches, aiming to design unique functionalities and enhance inherent properties, providing notable advantages [15-17]. However, applying band alignment strategies to manipulate the electronic and optical properties of layered metallic materials remains less understood, presenting an open area for the exploration of the two-dimensional materials, especially for non-centrosymmetric semimetals like TaIrTe$_4$.

With this perspective, we designed a dual-probe devices (Fig. 1a) of TaIrTe$_4$-based heterostructures with varying types of band alignment, and carried out a comprehensive density functional theory (DFT) and nonequilibrium Green's function (NEGF) analysis of their electronic and quantum transport properties. The pristine TaIrTe$_4$ monolayer manifest semimetal-like dispersions near Fermi level (Fig. 1b), while Weyl points of different chiralities are symmetrically distributed on both sides along the high symmetry line. Three monolayer materials with distinct band gaps and band edge positions (Fig. 1c) are used to create heterostructures with TaIrTe$_4$: TaIrTe$_4$/WTe$_2$, TaIrTe$_4$/MoTe$_2$, and TaIrTe$_4$/$h$-BN. Among them, 1T'-WTe$_2$ is another orthorhombic TMD type-II WSM with a similar structure and band edge position [18]; MoTe$_2$ is a TMD semiconductor with an optical band around 1.1 eV [19], providing a moderate energy window for TaIrTe$_4$'s band edge states; $h$-BN, a wide bandgap material (~6.9 eV), is widely used as encapsulation material, believed to protect the band-edge states from the external chemical environment influences; their band structures are given in Fig. S1. Two main aspects of the band alignment strategy are investigated: electrical interlayer doping via charge transfer and built-in electric field due to chemical potential differences. Our findings indicate that while all heterogeneous integrations can enhance the photoresponse, the underlying mechanisms are diverse.

The rest of the paper is structured as follows: the next section presents the calculation method. Sections III details the structural characteristics of the three heterostructures and explores their electronic properties under different band alignment strategies, which results in distinct interlayer interactions and charge transfer. The anisotropy and enhancement of photoresponse thus are investigated and understood. Finally, the last section includes a brief discussion and conclusion.

**Computational details**

The lattice structure was optimized by using the meta-generalized gradient approximation (meta-GGA) within the strongly constrained and appropriately normed (SCAN) functional plus revised Vydrov-van Voorhis nonlocal correlation rVV10 [20], as implemented in Vienna ab initio simulation package (VASP) [21]. The SCAN+rVV10 functional has demonstrated reliable accuracy in predicting the structural properties of a variety of layered materials [20]. To minimize interactions between periodic adjacent layers in few-layer systems, a vacuum layer exceeding 20 Å was applied. An energy cutoff of 500 eV and a 6×4×1 k-point grid were used for the heterostructure calculations. Transport calculations were performed by using the quantum transport software NANODCAL [22] within the nonequilibrium Green's function (NEGF) formalism [23,24]. The DZP atomic orbital basis is used to expand all the physical quantities with an energy cutoff of 80 Hartree; the exchange-correlation functional is within generalized gradient approximation (GGA) at the Perdew, Burke and Ernzerhof (PBE) level [25]; atomic cores were represented by standard norm-conserving nonlocal pseudopotentials; a Fermi smearing of 0.1 eV and a 10×1 k points were used.

**Results and discussion**

**A. Structural properties of TaIrTe$_4$-based heterostructures**

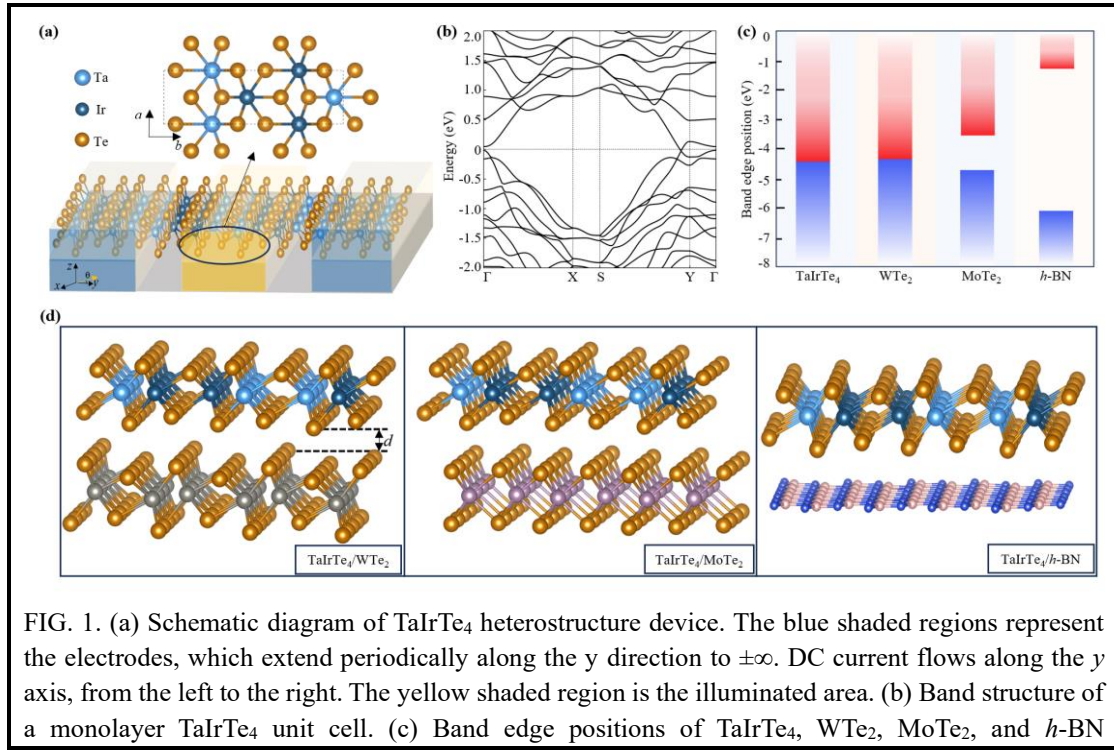

FIG. 1. (a) Schematic diagram of TaIrTe$_4$ heterostructure device. The blue shaded regions represent the electrodes, which extend periodically along the y direction to ±∞. DC current flows along the *y* axis, from the left to the right. The yellow shaded region is the illuminated area. (b) Band structure of a monolayer TaIrTe$_4$ unit cell. (c) Band edge positions of TaIrTe$_4$, WTe$_2$, MoTe$_2$, and *h*-BN

monolayers. (d) Schematic diagrams of the TaIrTe$_4$/WTe$_2$, TaIrTe$_4$/MoTe$_2$, and TaIrTe$_4$/$h$-BN heterostructures.

Bulk TaIrTe$_4$ crystallizes in a layered structure with space group Pmn2$_1$, similar to other T$_d$ TMD structures, manifesting zigzag-like distortions along both $a$ and $b$ axes. Its monolayer comprises three atomic layers (Fig. 1), with Ta or Ir atoms adopting a tetrahedral coordination and Te atoms forming three covalent bonds with them. Theoretically calculated lattice constants for the TaIrTe$_4$ monolayer unit cell are $a$ = 3.77 Å, $b$ = 12.47 Å. The geometric structures of TaIrTe$_4$/WTe$_2$, TaIrTe$_4$/MoTe$_2$, and TaIrTe$_4$/$h$-BN heterostructures are shown in Fig. 1d. Table I compares the optimized lattice constants and interlayer distance $d$ (defined in Fig. 1d) for the three heterostructures. Relative to the pristine TaIrTe$_4$ monolayer, the area mismatch rates for TaIrTe$_4$/WTe$_2$, TaIrTe$_4$/MoTe$_2$, and TaIrTe$_4$/$h$-BN membranes are -4.1%, -4.9% and 3.7%, respectively. Additionally, compared to the interlayer distance of 2.51 Å in bulk TaIrTe$_4$, the interlayer distances of the heterostructures are larger, indicating the weaker interlayer coupling on structural properties.

Table I. Lattice constants of the monolayer TaIrTe$_4$, and the three heterostructures: TaIrTe$_4$/WTe$_2$, TaIrTe$_4$/MoTe$_2$, and TaIrTe$_4$/$h$-BN. $d$ is the interlayer distance. The Mismatch rate is defined as the change in area between the heterostructure and pristine monolayer TaIrTe$_4$.

|  | TaIrTe$_4$ | TaIrTe$_4$/WTe$_2$ | TaIrTe$_4$/MoTe$_2$ | TaIrTe$_4$/$h$-BN |
|---|---|---|---|---|
| $a$ (Å) | 7.54 | 7.21 | 7.27 | 7.53 |
| $b$ (Å) | 12.48 | 12.52 | 12.31 | 12.94 |
| $d$ (Å) |  | 2.75 | 3.13 | 3.39 |
| Mismatch |  | -4.1% | -4.9% | 3.7% |

## B. Band structure, charge transfer and built-in electric field

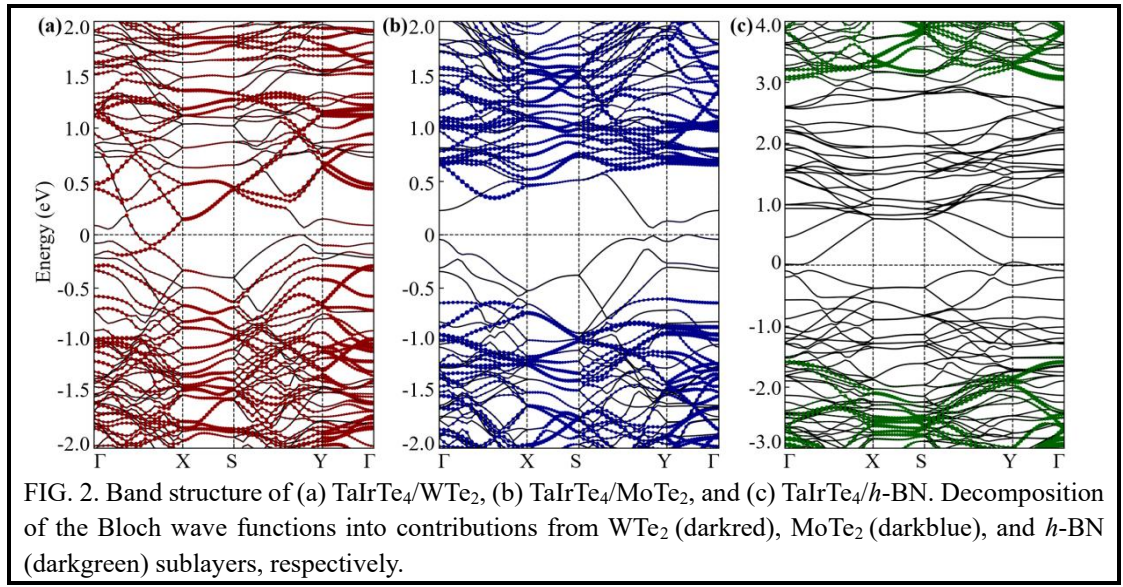

FIG. 2. Band structure of (a) TaIrTe$_4$/WTe$_2$, (b) TaIrTe$_4$/MoTe$_2$, and (c) TaIrTe$_4$/$h$-BN. Decomposition of the Bloch wave functions into contributions from WTe$_2$ (darkred), MoTe$_2$ (darkblue), and $h$-BN (darkgreen) sublayers, respectively.

Figure 2 depicts the projected band structures of the three heterostructures, where contributions from WTe$_2$, MoTe$_2$, and $h$-BN sublayers are colored in darkred, darkblue and darkgreen dots, respectively. In TaIrTe$_4$/WTe$_2$, due to their similarity on both metallicity and band edge positions, the states of WTe$_2$ and TaIrTe$_4$ are highly mixed in a broad energy window. In contrast, the band edge states in TaIrTe$_4$/MoTe$_2$ and TaIrTe$_4$/*h*-BN predominantly arise from TaIrTe$_4$ sublayer within the energy windows defined by the band gaps of MoTe$_2$ and *h*-BN. The interlayer coupling results in slight deviations in band offsets compared to isolated monolayers (detailed values are listed in Table S1). For example, the ionization energy offset between isolated TaIrTe$_4$ monolayer and MoTe$_2$ monolayer is 0.50 eV (Table S1 and Fig. 1c), which increase to 0.70 eV in TaIrTe$_4$/MoTe$_2$ heterostructure as shown in Fig. 2b. Instead, the energy affinity offset between isolated TaIrTe$_4$ monolayer and MoTe$_2$ monolayer is 0.67 eV (Table S1), which reduce to 0.30 eV in TaIrTe$_4$/MoTe$_2$ heterostructure (Fig. 2b), indicating that the bottom conduction bands of MoTe$_2$ sublayer may facilitate the transport and photoresponse in TaIrTe$_4$.

The band edge dispersions of TaIrTe$_4$, particularly near the Fermi level, undergo substantial reconstruction due to the heterogeneous integration. Interlayer strain arising from the lattice mismatch between bilayer (defined in Table I) significantly impacts those dispersion changes. To verify this point, we plot the band structures of the TaIrTe$_4$ monolayer by using its equilibrium lattice parameters (strain free) and the strained parameters matching those of the three heterostructures for comparison [see Fig. S2 and Fig. S3]. We observe that the band edge dispersions of TaIrTe$_4$ within heterostructures inherits most changes of the strained TaIrTe$_4$ monolayer, while hybridization and interlayer interaction further enhance these dispersion modifications. Clearly, the strong hybridization and interweaving of band-edge states determine the broadband electronic properties of TaIrTe$_4$/WTe$_2$. Then, interlayer interaction results in the valence-band splitting and shifts the relative positions of the conduction bands of the TaIrTe$_4$/MoTe$_2$ heterostructure. For the TaIrTe$_4$/*h*-BN heterostructure, the interlayer interaction is weak, but the band splitting near the Fermi level still undergoes a slight modification. These band edge changes will influence transport and photoresponse properties, underscoring the importance of a comprehensive understanding.

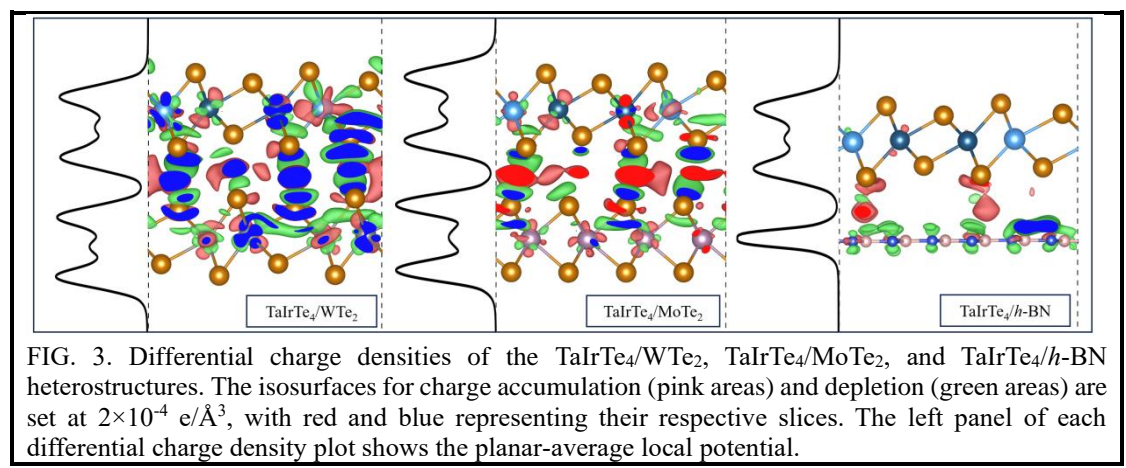

FIG. 3. Differential charge densities of the TaIrTe$_4$/WTe$_2$, TaIrTe$_4$/MoTe$_2$, and TaIrTe$_4$/*h*-BN heterostructures. The isosurfaces for charge accumulation (pink areas) and depletion (green areas) are set at $2\times10^{-4}$ e/Å$^3$, with red and blue representing their respective slices. The left panel of each differential charge density plot shows the planar-average local potential.

The interlayer interactions primarily involve the charge transfer and built-in electric field, which arise from differences of the chemical potential. The differential charge density and planar-average local potential are depicted in Fig. 3. The charge accumulation (pink) and depletion (green) intuitively illustrate the spatial charge transfer. In the TaIrTe$_4$/WTe$_2$ system, the high hybridization and metallic band interweaving make the interlayer charge transfer and built-in electric fields less meaningful to discuss. Similar local potential (left panel in Fig. 3) in the upper and lower sublayers promotes the interlayer charge mixing, with the overall trend being charge accumulation in the intermediate region. This accumulation also occurs in the TaIrTe$_4$/MoTe$_2$ and TaIrTe$_4$/$h$-BN systems, indicating that the interlayer charger transfer effect is highly localized regardless of the band alignment. Due to the lower local potential of MoTe$_2$ sublayer relative to TaIrTe$_4$, a built-in electric field (higher in the upper layers and lower in the lower layers) forms in the TaIrTe$_4$/MoTe$_2$ heterostructure, resulting in the electron loss in the MoTe$_2$ sublayer and electron gain in the TaIrTe$_4$ sublayer. A similar but much smaller charge transfer trend is observed in the TaIrTe$_4$/$h$-BN system, forming the basis for the variations in the band dispersion and offset.

## C. Enhanced broadband photoresponse in heterostructures

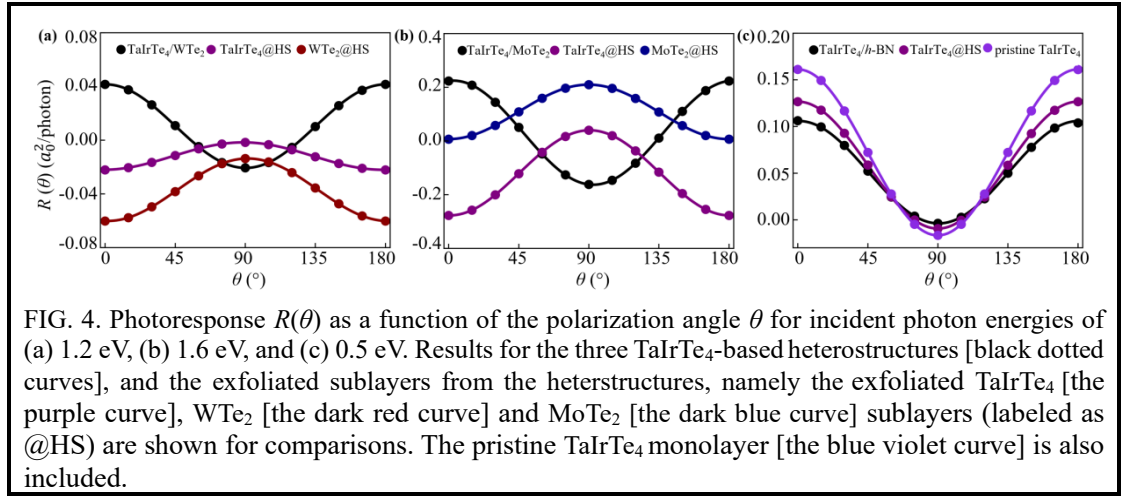

FIG. 4. Photoresponse $R(\theta)$ as a function of the polarization angle $\theta$ for incident photon energies of (a) 1.2 eV, (b) 1.6 eV, and (c) 0.5 eV. Results for the three TaIrTe$_4$-based heterostructures [black dotted curves], and the exfoliated sublayers from the heterstructures, namely the exfoliated TaIrTe$_4$ [the purple curve], WTe$_2$ [the dark red curve] and MoTe$_2$ [the dark blue curve] sublayers (labeled as @HS) are shown for comparisons. The pristine TaIrTe$_4$ monolayer [the blue violet curve] is also included.

Next, we investigate the photoresponse of TaIrTe$_4$-based heterostructures. A two-probe photodetector device was constructed, comprising a central region sandwiched by left and right electrodes, as shown in Fig. 1a. The electrodes extend to $y = \pm\infty$, where photocurrent is collected. The central region consists of unit cells arranged in a three-period lattice along the $b$ direction, with linear polarized light is shined normally on the center unit cell. We consider the photon energies from 0.1 to 2 eV, which covers the infrared region and part of the visible spectrum. Due to the lack of inversion symmetry, the TaIrTe$_4$-based device exhibits the photogalvanic effect (PGE), allowing a photocurrent generation without the bias voltage, thus achieving a self-powering phenomenon. Both the intensity and direction of photocurrent depend on the polarization angle of the incident light. For linearly polarized light, the photocurrent follows the form $R(\theta) = A \times \cos(2\theta) + R_0$, where $\theta$ is the polarization angle with respect to the transport direction, and $A$ depends on the device symmetry and photon energy. Similar to the results in the linear PGE in monolayer TaIrTe$_4$ [26] and WTe$_2$ [27], the maximum photoresponse intensity $|R|_{max}$ for the three heterostructures at either $\theta = 0°$

(or 180°) or $\theta = 90°$, depends on the photon energy, as shown in Fig. 4. For most of incident photon energies, the polarization angle of $|R|_{max}$ aligns with the transport direction ($\theta = 0°$ or $\theta = 180°$) for all monolayer and bilayer devices. However, in the case at 1.6 eV in MoTe$_2$@HS system, the polarization angle of $|R|_{max}$ is perpendicular to the transport direction.

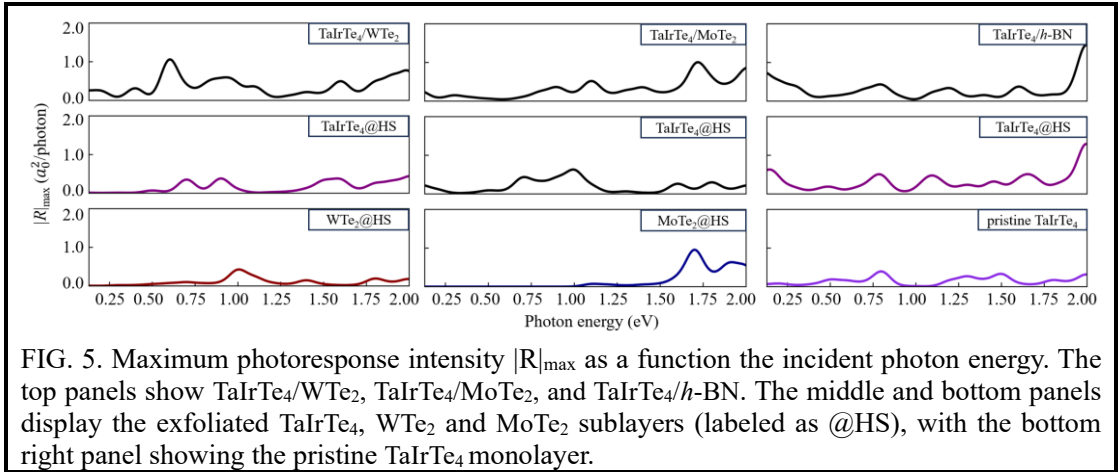

FIG. 5. Maximum photoresponse intensity $|R|_{max}$ as a function the incident photon energy. The top panels show TaIrTe$_4$/WTe$_2$, TaIrTe$_4$/MoTe$_2$, and TaIrTe$_4$/$h$-BN. The middle and bottom panels display the exfoliated TaIrTe$_4$, WTe$_2$ and MoTe$_2$ sublayers (labeled as @HS), with the bottom right panel showing the pristine TaIrTe$_4$ monolayer.

Figure 5 compares the photoresponse spectra of the three heterostructures and their exfoliated sublayers, and the pristine TaIrTe$_4$ monolayer. As expected, the heterogeneous integration significantly modifies the photoresponse of TaIrTe$_4$. Each heterostructure, utilizing a distinct band alignment strategy, enhances the overall photoresponse, though the underlying mechanisms are not exactly the same. In the TaIrTe$_4$/WTe$_2$ heterostructure, the maximum photoresponse occurs around 0.6 eV, considerably stronger than those of the exfoliated TaIrTe$_4$ and WTe$_2$ sublayers. To clearly understand these photocurrent peaks, we plot the device density of states (DOS) in Fig. 6. The photoresponse peak near 0.6 eV corresponds to the optical transition between the hole DOS peak at -0.2 eV and the electron DOS peak at 0.4 eV. Due to the band hybridization and overlap, these DOS peaks particularly the electron DOS peak of the TaIrTe$_4$/WTe$_2$ heterostructure in the far infrared region (0-0.5 eV) are strongly enhanced. Form the projected band structures (Fig. 2a), the hole DOS peak at -0.2 eV mainly originates from the TaIrTe$_4$ sublayer, while the electron DOS peak at 0.4 eV is primarily derived from WTe$_2$ sublayer. Thus, the maximum photoresponse in the TaIrTe$_4$/WTe$_2$ heterostructure largely results from interlayer optical transitions. This interlayer transition mechanism similarly enhances the far-infrared response in TaIrTe$_4$/WTe$_2$ and TaIrTe$_4$/MoTe$_2$ heterostructures.

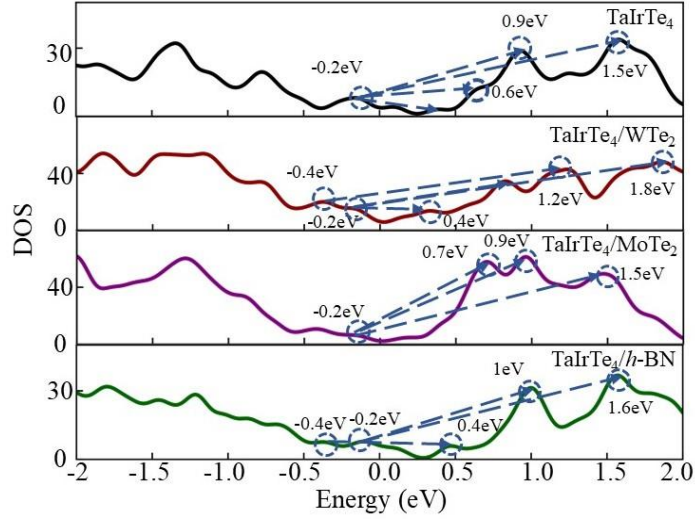

FIG. 6. Device density of states (DOS) for the photodetector based on the pristine TaIrTe$_4$ monolayer and the three heterostructures. Blue arrows indicate the DOS peaks corresponding to the optical transition responsible for the photoresponse peak.

Furthermore, the staggered optical responses between sublayers enhance the broadband response properties of the heterogeneously integrated device. For instance, the TaIrTe$_4$ sublayer dominates the infrared response in the TaIrTe$_4$/MoTe$_2$ heterostructure, while the MoTe$_2$ sublayer dominates in the visible region. In the TaIrTe$_4$/h-BN heterostructure, the photoresponse enhancement relative to the pristine TaIrTe$_4$ monolayer is primarily due to the in-plane strain, as indicated in the right panel of Fig. 5. The tensile strain effect of h-BN significantly boosts the photocurrent, which is more than the effects given by the compressive strains in the WTe$_2$ and MoTe$_2$ heterostructures. The TaIrTe$_4$@HS in TaIrTe$_4$/h-BN system shows a substantial enhancement at both spectrum ends (the infrared and 2.0 eV in the visible region). Additionally, the TaIrTe$_4$/h-BN heterostructure further improves the overall photoresponse due to the lower symmetry in the heterostructured device. Notably, although the heterogeneous integration enhances the overall photoresponse, it is not beneficial across all photon energies. For example, at around 0.7 eV in the TaIrTe$_4$/MoTe$_2$ system, the maximum photoresponse intensity of TaIrTe$_4$@HS is higher than that of TaIrTe$_4$/MoTe$_2$. At this photon energy, the heterogeneous integration degrades the device's photoresponse performance. This reduction is attributed to interlayer coupling effects that alter specific band dispersions. Additionally, the directions of generated photocurrent in the two sublayers may oppose each other at certain photon energies. For the example shown in Fig. 4b, the photocurrent direction of MoTe$_2$@HS sublayer is oriented from the central region towards the electrode, while the TaIrTe$_4$@HS sublayer is from the electrode towards the central region. This explains the total photocurrent in TaIrTe$_4$/MoTe$_2$ at 1.6 eV is slightly smaller than that of TaIrTe$_4$@HS.

## II. Summary

We conducted comprehensive DFT and NEGF calculations on TaIrTe$_4$/WTe$_2$,

TaIrTe$_4$/MoTe$_2$ and TaIrTe$_4$/$h$-BN heterostructures. These TaIrTe$_4$-based systems exhibit distinct band alignments, with the heterogeneous integration significantly modifying their electronic and optical properties. We carefully analyzed the effects of interlayer interactions including the hybridization, charge transfer, and built-in electric fields on the electronic structures. The $h$-BN sublayer induces a tensile strain, particularly enhancing the photoresponse in both the infrared region and visible spectrum around 2.0 eV. While WTe$_2$ and MoTe$_2$ sublayers creat a compressive strain, promoting a broadband response. Additionally, these self-powered photodetectors were demonstrated with enhanced photoresponses explained in terms of the mechanisms such as interlayer transitions, band overlap, and symmetry lowering induced by the heterogeneous integration.

**Acknowledgments**

Work at UCAS was supported by National Natural Science Foundation of China under Grants No. 12347115. Work at HDU was supported by National Natural Science Foundation of China (Grant No. 11574067, 12104118), the Natural Science Foundation of Zhejiang Province (Grant No. LTGS24E060005), and the Foundation of Hangzhou Dianzi University (Grants No. KYS075624288).

# References

[1] J. Tang *et al.*, *Dual quantum spin hall insulator by density-tuned correlations in tairte4,* Nature **628**, 515 (2024).

[2] K. Koepernik, D. Kasinathan, D. Efremov, S. Khim, S. Borisenko, B. Büchner, and J. van den Brink, *Tairte 4: A ternary type-ii weyl semimetal,* Physical Review B **93**, 201101 (2016).

[3] J. Lai, Y. Liu, J. Ma, X. Zhuo, Y. Peng, W. Lu, Z. Liu, J. Chen, and D. Sun, *Broadband anisotropic photoresponse of the "hydrogen atom" version type-ii weyl semimetal candidate tairte4,* ACS nano **12**, 4055 (2018).

[4] X. Han *et al.*, *A polarization-sensitive self-powered photodetector based on a p-wse2/tairte4/n-mos2 van der waals heterojunction,* ACS Applied Materials & Interfaces **13**, 61544 (2021).

[5] J. Lai *et al.*, *Direct light orbital angular momentum detection in mid-infrared based on the type-ii weyl semimetal tairte4,* Advanced Materials **34**, 2201229 (2022).

[6] J. Ma *et al.*, *Nonlinear photoresponse of type-ii weyl semimetals,* Nature materials **18**, 476 (2019).

[7] X. Zhuo, J. Lai, P. Yu, Z. Yu, J. Ma, W. Lu, M. Liu, Z. Liu, and D. Sun, *Dynamical evolution of anisotropic response of type-ii weyl semimetal tairte4 under ultrafast photoexcitation,* Light: Science & Applications **10**, 101 (2021).

[8] X. Dong *et al.*, *Observation of topological edge states at the step edges on the surface of type-ii weyl semimetal tairte4,* ACS nano **13**, 9571 (2019).

[9] I. Belopolski *et al.*, *Signatures of a time-reversal symmetric weyl semimetal with only four weyl points,* Nature communications **8**, 942 (2017).

[10] D. Kumar, C.-H. Hsu, R. Sharma, T.-R. Chang, P. Yu, J. Wang, G. Eda, G. Liang, and H. Yang, *Room-temperature nonlinear hall effect and wireless radiofrequency rectification in weyl semimetal tairte$_4$,* Nature Nanotechnology **16**, 421 (2021).

[11] Y. Liu *et al.*, *Raman signatures of broken inversion symmetry and in-plane anisotropy in type-ii weyl semimetal candidate tairte4,* Advanced Materials **30**, 1706402 (2018).

[12] R. Liu, F. Wang, L. Liu, X. He, J. Chen, Y. Li, and T. Zhai, *Band alignment engineering in two-dimensional transition metal dichalcogenide-based heterostructures for photodetectors,* Small Structures **2**, 2000136 (2021).

[13] S. S. Lo, T. Mirkovic, C. H. Chuang, C. Burda, and G. D. Scholes, *Emergent properties resulting from type-ii band alignment in semiconductor nanoheterostructures,* Advanced Materials **23**, 180 (2011).

[14] Y. Chen *et al.*, *Momentum-matching and band-alignment van der waals heterostructures for high-efficiency infrared photodetection,* Science Advances **8**, eabq1781 (2022).

[15] K. Moore, P. Dawson, and C. Foxon, *Effects of electronic coupling on the band alignment of thin gaas/alas quantum-well structures,* Physical Review B **38**, 3368 (1988).

[16] M. Thewalt, D. Harrison, C. Reinhart, J. Wolk, and H. Lafontaine, *Type ii band alignment in si 1− x ge x/si (001) quantum wells: The ubiquitous type i luminescence results from band bending,* Physical Review Letters **79**, 269 (1997).

[17] Y.-H. Kuo, Y. K. Lee, Y. Ge, S. Ren, J. E. Roth, T. I. Kamins, D. A. Miller, and J. S. Harris, *Strong quantum-confined stark effect in germanium quantum-well structures on silicon,* Nature **437**, 1334 (2005).

[18] Z. Fei *et al.*, *Edge conduction in monolayer wte2,* Nature Physics **13**, 677 (2017).

[19] C. Ruppert, B. Aslan, and T. F. Heinz, *Optical properties and band gap of single-and few-layer mote2 crystals,* Nano letters **14**, 6231 (2014).

[20] H. Peng, Z.-H. Yang, J. P. Perdew, and J. Sun, *Versatile van der waals density functional based on*

*a meta-generalized gradient approximation,* Physical Review X **6**, 041005 (2016).

[21] G. Kresse and J. Furthmuller, *Efficient iterative schemes for ab initio total-energy calculations using a plane-wave basis set,* Physical review. B, Condensed matter **54**, 11169 (1996).

[22] Y. Xie, M. Chen, Z. Wu, Y. Hu, Y. Wang, J. Wang, and H. Guo, *Two-dimensional photogalvanic spin-battery,* Physical Review Applied **10**, 034005 (2018).

[23] L. E. J. J. o. a. p. Henrickson, *Nonequilibrium photocurrent modeling in resonant tunneling photodetectors,* **91**, 6273 (2002).

[24] Y. Xie, L. Zhang, Y. Zhu, L. Liu, and H. J. N. Guo, *Photogalvanic effect in monolayer black phosphorus,* **26**, 455202 (2015).

[25] J. P. Perdew, K. Burke, and M. J. P. r. l. Ernzerhof, *Generalized gradient approximation made simple,* **77**, 3865 (1996).

[26] Y. Ding, X. Wang, L. Liao, X. Cheng, J. Zhang, Y. Wang, H. Ying, and Y. Li, *Strain modulation of photocurrent in weyl semimetal tairte 4,* Optics Letters **47**, 4881 (2022).

[27] Z. Xu, B. Luo, Y. Chen, X. Li, Z. Chen, Q. Yuan, and X. Xiao, *High sensitivity and anisotropic broadband photoresponse of td-wte2,* Physics Letters A **389**, 127093 (2021).